\documentclass{elsart}
\usepackage{graphicx}
\usepackage{epsfig}
\usepackage{amssymb}

\begin{document}

\begin{frontmatter}


\title{Variation of hadron masses in nuclear matter in the relativistic
Hartree approximation}

\author[a1]{H. Shen}            \ead{songtc@public.tpt.tj.cn}
\author[a2]{S. Tamenaga}        \ead{stame@rcnp.osaka-u.ac.jp}
\author[a2]{H. Toki}            \ead{toki@rcnp.osaka-u.ac.jp}

\address[a1]{Department of physics, Nankai University, Tianjin 300071, China}
\address[a2]{Research Center for Nuclear Physics (RCNP), Osaka University,
             Ibaraki, Osaka 567-0047, Japan}

\begin{abstract}
We study the modification of hadron masses due to the
vacuum polarization using the chiral sigma model, which is
extended to generate the $\omega$ meson mass by the sigma
condensation in the vacuum in the same way as the nucleon mass.
The results obtained in the chiral sigma model are compared with
those obtained in the Walecka model which includes $\sigma$ and
$\omega$ mesons in a non-chiral fashion. It is shown that both the
nucleon mass and the $\omega$ meson mass decrease in nuclear
medium, while the $\sigma$ meson mass increases at finite density
in the chiral sigma model.
\end{abstract}

\end{frontmatter}

\section{Introduction}

One of the most interesting topics in nuclear physics is to study
how the hadron properties are altered as the environment changes.
In particular, the medium modification of hadron masses has
attracted a lot of attention both experimentally and
theoretically. The observation of enhanced dilepton production
from relativistic heavy ion collision experiments~\cite{dilepton}
could be due to a reduction in the vector meson masses in the
medium. Brown and Rho suggested the hypothesis that the vector
meson masses drop in nuclear medium according to a simple scaling
law~\cite{brownrho}. There are many theoretical efforts made to
understand the behavior of hadrons in dense matter, including the
various QCD-based methods like QCD sum rules~\cite{sumrule} and
phenomenological nuclear models such as relativistic mean field
approach~\cite{walecka} and quark meson coupling model~\cite{qmc}.

The relativistic mean field theory (RMF) has been developed and
widely applied to a great variety of problems in nuclear
physics~\cite{walecka}. The original version of the RMF theory
proposed by Walecka (Walecka model) consists of baryons
interacting with each other via the exchange of $\sigma$ and
$\omega$ mesons~\cite{walecka}. This model and its later
variations can be solved in the mean field approximation, by
replacing the meson field operators with their classical
expectation values. Since the Walecka model is renormalizable,
there is a standard procedure for renormalization of the vacuum
polarization contribution carried out by adding required
counterterms to the original Lagrangian and subtracting purely
vacuum expectation values. It has been discussed that the Walecka
model with the vacuum polarization effects taken into account
could reproduce reasonably well the saturation properties of
nuclear matter and the ground state properties of finite
nuclei~\cite{walecka}. However, the Walecka model does not respect
the chiral symmetry, which is known to be a very important feature
in hadron physics.

The chiral symmetry can be described nicely in the linear sigma model
introduced by Gell-Mann and Levy~\cite{gellmann}, which has been used
for various phenomena in hadron physics.
It is very natural to use the chiral sigma model for the description
of nuclear matter and finite nuclei in the relativistic mean field
approximation~\cite{walecka}.
It was found that the use of the chiral sigma model in its original form
was not satisfactory for the description of nuclear matter.
Boguta introduced a dynamical generation of the $\omega$ meson mass
in the same way as the nucleon mass,
so that a saturating equation of state for nuclear matter could
be obtained in the chiral sigma model~\cite{boguta}.
This chiral sigma model was applied to study finite nuclei
by several groups~\cite{savushkin,ogawa}.

Many authors have studied the medium modification of hadron masses
using some non-chiral models~\cite{suzuki2,hatsuda1,jean,saito}.
It has been pointed out that the polarization of the Dirac sea is
the most important reason for the reduction of vector meson masses
in medium~\cite{hatsuda1}. In this paper, we would like to
investigate the variation of hadron masses due to vacuum
polarization within the chiral sigma model, and compare with the
results obtained in the Walecka model. In section~\ref{sec: rmf},
we briefly recapitulate the non-chiral Walecka model and the
chiral sigma model, also discuss the procedure for renormalization
in these models. In section~\ref{sec: num}, we explain the model
parameters, and show the numerical results. Section~\ref{sec: sum}
is devoted to the summary of this paper.

\section{Formalism}
\label{sec: rmf}

In this section, we briefly recapitulate the effective Lagrangian and the
renormalization procedure in the Walecka model and in the chiral sigma model.
The details regarding the renormalization procedure can be found in earlier
references~\cite{chin,matsui,suzuki1}.

The Lagrangian density of the Walecka model is well known, which involves
an explicit description of nucleon and meson degrees of freedom~\cite{walecka}.
The Lagrangian density in the Walecka model is given as
\begin{eqnarray}
{\mathcal L} &=&
\bar{\Psi}\left(i\gamma_{\mu}\partial^{\mu}-M-g_{\sigma}\sigma
              -g_{\omega}\gamma_{\mu}\omega^{\mu}\right)\Psi \\ \nonumber
 & &+\frac{1}{2}\partial_{\mu}\sigma\partial^{\mu}\sigma
    -\frac{1}{2}m_{\sigma}^2\sigma^2
    -\frac{1}{4}W_{\mu\nu}W^{\mu\nu}+\frac{1}{2}m_{\omega}^2\omega_{\mu}\omega^{\mu}
 + \delta {\mathcal L},
\label{eq:lw}
\end{eqnarray}
where $\Psi$, $\sigma$ and $\omega$ are the fields for the
nucleon, $\sigma$ and $\omega$ mesons with physical masses $M$,
$m_{\sigma}$ and $m_{\omega}$, respectively, and $W^{\mu\nu}
\equiv \partial^{\mu}\omega^{\nu}-\partial^{\nu}\omega^{\mu}$. The
term $\delta {\mathcal L} = \delta {\mathcal L}_\sigma + \delta
{\mathcal L}_\omega$ contains renormalization counterterms, which
are introduced to remove the divergences in the loop calculations
within the framework of the relativistic Hartree
approximation~\cite{walecka}. The renormalization procedure in the
Walecka model has been extensively discussed in
Refs.~\cite{chin,matsui,suzuki1}. Here, we adopt the subtraction
scheme given in Ref.~\cite{suzuki1}. The renormalization
counterterms in the Walecka model can be written as
\begin{eqnarray}
\delta {\mathcal L}_\sigma &=&
\alpha_{1}\sigma+\frac{1}{2!}\alpha_{2}\sigma ^2
                          +\frac{1}{3!}\alpha_{3} \sigma ^3+\frac{1}{4!}\alpha_{4} \sigma^4
                          +\frac{1}{2}\zeta_{\sigma}\partial_{\mu}\sigma\partial^{\mu}\sigma,
\\
\delta {\mathcal L}_\omega
&=&-\frac{1}{4}\zeta_{\omega}W_{\mu\nu}W^{\mu\nu}
                          -\frac{1}{2}\delta m_{\omega}^2\omega_{\mu}\omega^{\mu}.
\end{eqnarray}
The coefficients are specified by imposing appropriate
renormalization conditions. First of all, $\alpha_1$ must
completely cancel the loop contribution to ensure the stability of
the vacuum. The coefficients $\alpha_2$ and $\zeta_\sigma$ can be
determined by requiring $\Pi^R_{\sigma}|_{M^*=M,q^2=m^2_\sigma}=0$
and $\frac{\partial \Pi^R_{\sigma}}{\partial
q^2}|_{M^*=M,q^2=m^2_\sigma}=0$. For $\alpha_3$ and $\alpha_4$, we
adopt the usual conditions used in Ref.~\cite{walecka}. The
coefficients $\delta m_{\omega}^2$ and $\zeta_\omega$ can be
determined by imposing $D_\omega |_{M^*=M,q^2=m^2_\omega} = 0$ and
$\frac{\partial D_\omega}{\partial
q^2}|_{M^*=M,q^2=m^2_\omega}=1$, where $D_\omega =
q^2-m_\omega^2+\delta m_\omega^2 -q^2 \Pi^R_{\omega}$. The
explicit expressions for the renormalized meson self-energies in
the Walecka model, $\Pi^R_{\sigma}$ and $\Pi^R_{\omega}$, can be
found in Ref.~\cite{suzuki1}. After carrying out the
renormalization procedure, we study the effective masses of
$\sigma$ and $\omega$ mesons, $m^*_\sigma$ and $m^*_\omega$, which
can be obtained by searching for the zeros of the inverse
propagators,
\begin{eqnarray}
D_\sigma (M^*,q^2={m_\sigma^*}^2) &=& q^2-m_\sigma^2-\Pi^R_{\sigma}(M^*,q^2)=0
\label{eq:msw} \\
D_\omega (M^*,q^2={m_\omega^*}^2) &=&
                  q^2-m_\omega^2+\delta m_\omega^2 -q^2 \Pi^R_{\omega}(M^*,q^2)=0
\label{eq:mww}
\end{eqnarray}

We now turn to the chiral sigma model used in Ref.~\cite{ogawa}.
The Lagrangian density of the chiral sigma model is written as
\begin{eqnarray}
{\mathcal L} & = &
\bar{\Psi}\left[i\gamma_{\mu}\partial^{\mu}
                - g_{\sigma}\left(\sigma + i\gamma_{5}{\vec\tau}\cdot{\vec\pi}\right)
                - g_{\omega}\gamma_{\mu}\omega^{\mu}\right]\Psi \\ \nonumber
 & & +\frac{1}{2} \partial_{\mu}\sigma \partial^{\mu}\sigma
                 + \frac{1}{2} \partial_{\mu}{\vec\pi} \partial^{\mu}{\vec\pi}
                 - \frac{\mu^{2}}{2}\left(\sigma^2 + {\vec\pi}^2\right)
                 - \frac{\lambda}{4}\left(\sigma^2 + {\vec\pi}^2\right)^2 \\ \nonumber
 & & -\frac{1}{4} W_{\mu\nu}W^{\mu\nu}
                 + \frac{1}{2}\widetilde{g}_{\omega}^2
                   \left(\sigma^2 + {\vec\pi}^2\right)\omega_{\mu}\omega^{\mu} \\ \nonumber
 & & +\varepsilon\sigma + \delta\mathcal{L}.
\label{eq:lsw}
\end{eqnarray}
Here $\Psi$, $\pi$, $\sigma$ and $\omega$ are the fields for the
nucleon, $\pi$, $\sigma$ and $\omega$ mesons. The $\omega$ meson
mass can be generated dynamically by the sigma condensation in the
vacuum in the same way as the nucleon mass\cite{boguta}. In the
Lagrangian, an explicit chiral symmetry breaking term
$\varepsilon\sigma$ has been involved, while the term
$\delta\mathcal{L}$ contains the renormalization counterterms. To
realize the chiral symmetry in the Nambu-Goldstone mode, a nonzero
vacuum expectation value of the $\sigma$ field, $\langle \sigma
\rangle = \sigma_0$, is obtained by minimizing the meson effective
potential. We now define the new fluctuation field
$\varphi=\sigma-\sigma_0$, the above Lagrangian is rewritten as
\begin{eqnarray}
 {\mathcal L}& = & \bar{\Psi}\left(i\gamma_{\mu}\partial^{\mu} - M
             - g_{\sigma}\varphi -g_{\sigma}i\gamma_{5}{\vec\tau}\cdot{\vec\pi}
             - g_{\omega}\gamma_{\mu}\omega^{\mu}\right)\Psi \\ \nonumber
       &   & + \frac{1}{2}  \partial_{\mu}{\vec\pi} \partial^{\mu}{\vec\pi}
             + \frac{1}{2}  \partial_{\mu}\varphi \partial^{\mu}\varphi
             - \frac{1}{4}  W_{\mu\nu}W^{\mu\nu} \\ \nonumber
       &   & - \frac{1}{2}m_\pi^2 {\vec\pi}^2
             - \frac{1}{2}m_\sigma^2 \varphi^2
             + \frac{1}{2} m_\omega^2 \omega_{\mu}\omega^{\mu} \\ \nonumber
       &   & - \lambda\sigma_0 \varphi^3 - \frac{\lambda}{4} \varphi^4
             - \frac{\lambda}{4}
               \left( 4\sigma_0 \varphi + 2\varphi^2 + {\vec\pi}^2\right){\vec\pi}^2 \\ \nonumber
       &   & + \frac{1}{2}\widetilde{g}_{\omega}^2
               \left( 2\sigma_0 \varphi + \varphi^2 + {\vec\pi}^2\right)
               \omega_{\mu}\omega^{\mu} \\ \nonumber
        &  & + \left(\varepsilon-\mu^2\sigma_0 - \lambda\sigma_0^3\right)\varphi
             + \delta {\mathcal L}.
\nonumber
\end{eqnarray}
Here we have dropped a non-essential c-number constant.
The energy minimum condition requires the term linear in $\varphi$ to be zero,
$\varepsilon-\mu^2\sigma_0 - \lambda\sigma_0^3=0$.
The physical masses are related with the parameters in the Lagrangian as
\begin{eqnarray}
 M          &=& g_\sigma \sigma_0 ,                                                   \\
 m_\pi^2    &=& \mu^2 +  \lambda\sigma_0^2 = \varepsilon / \sigma_0,   \label{eq:ms1} \\
 m_\sigma^2 &=& \mu^2 + 3\lambda\sigma_0^2 ,                           \label{eq:ms2} \\
 m_\omega^2 &=& \widetilde{g}_{\omega}^2 \sigma_0^2 .
\end{eqnarray}
The parameter $\varepsilon$, which is proportional to $m^2_\pi$,
represents the order of the explicit chiral symmetry breaking,
and the exact chiral limit can be obtained by setting $\varepsilon=0$.
To perform the renormalization for the chiral sigma model,
we need the counterterms of the form
\begin{eqnarray}
\delta {\mathcal L} = \delta {\mathcal L}_{\sigma\pi} + \delta
{\mathcal L}_\omega.
\end{eqnarray}
Here, $\delta {\mathcal L}_\omega$ is identical to the one in the
Walecka model. The term $\delta {\mathcal L}_{\sigma\pi}$ can be
written as
\begin{eqnarray}
\delta {\mathcal L}_{\sigma\pi} &=&
      \alpha_1 \varphi +\frac{1}{2!}\alpha_2 \varphi^2
     +\frac{1}{3!}\alpha_3 \varphi^3 + \frac{1}{4!}\alpha_4 \varphi^4\\ \nonumber
{}& &+\frac{1}{2!}\beta_2{\vec\pi}^2 + \frac{1}{2!}\beta_3\varphi{\vec\pi}^2
     +\frac{1}{2!2!}\beta_4\varphi^2{\vec\pi}^2 \\ \nonumber
{}& &+\frac{1}{2}\zeta_{\sigma}\partial_{\mu}\sigma\partial^{\mu}\sigma
     +\frac{1}{2}\zeta_{\pi}\partial_{\mu}{\vec\pi}\partial^{\mu}{\vec\pi}+...
\end{eqnarray}
In order to respect the chiral symmetry in the renormalization
procedure, $\delta {\mathcal L}_{\sigma\pi}$ should get back a
chiral symmetric form in the limit $\varepsilon\rightarrow 0$, as
discussed in Ref.~\cite{matsui}. Therefore, much of the
arbitrariness in the renormalization procedure is eliminated, and
the coefficients $\alpha_i$ and $\beta_i$ in the chiral sigma
model are related to each other. We again take the following
renormalization conditions,
$\Pi^R_{\sigma}|_{M^*=M,q^2=m^2_\sigma}=0$ and $\frac{\partial
\Pi^R_{\sigma}}{\partial q^2}|_{M^*=M,q^2=m^2_\sigma}=0$, to
specify the independent coefficients in the counterterms. Now we
do not need any extra renormalization conditions to specify the
coefficients $\alpha_3$ and $\alpha_4$, as done in the Walecka
model. After carrying out the renormalization procedure, we can
study the effective masses of $\sigma$ and $\omega$ mesons in
nuclear medium, $m^*_\sigma$ and $m^*_\omega$. The effective
masses are obtained by searching for the zeros of the inverse
propagators,
\begin{eqnarray}
\hspace*{-1.2cm}
D_\sigma (M^*,q^2={m_\sigma^*}^2) &=& q^2-m_\sigma^2
  -6\lambda f_\pi\varphi -3\lambda\varphi^2 +\widetilde{g}_{\omega}^2 \omega^2
  -\Pi^R_{\sigma}(M^*,q^2)=0
\label{eq:msc}  \\
\hspace*{-1.2cm}
D_\omega (M^*,q^2={m_\omega^*}^2) &=& q^2 - m_\omega^2 + \delta m_\omega^2
  -2\widetilde{g}_{\omega}^2 f_\pi\varphi -\widetilde{g}_{\omega}^2 \varphi^2
  -q^2 \Pi^R_{\omega}(M^*,q^2)=0
\label{eq:mwc}
\end{eqnarray}
where $\Pi^R_{\omega}$ is the same as that in the Walecka model,
but $\Pi^R_{\sigma}$ in the chiral sigma model is different from the one
in the Walecka model due to the changes of the coefficients $\alpha_3$ and $\alpha_4$,
which can be written as
\begin{eqnarray}
\left(\Pi^R_{\sigma}\right)_{chiral}&=&
      \left(\Pi^R_{\sigma}\right)_{Walecka}
      -\Delta\alpha_3\frac{M^*-M}{g_\sigma}
      -\frac{\Delta\alpha_4}{2}\left(\frac{M^*-M}{g_\sigma}\right)^2.
\end{eqnarray}
$\Delta\alpha_3$ and $\Delta\alpha_4$ are the differences between the coefficients
in the two models, and we find
\begin{eqnarray}
\Delta\alpha_3&=& \frac{g_{\sigma}^3M}{\pi^2}
      \left\{2-\frac{3m_\sigma^2}{4 M^2}
              -\frac{9}{2}\int^1_0dxln\left[1-\frac{m_\sigma^2}{M^2}x(1-x)\right]\right\}\\
\Delta\alpha_4&=& \frac{g_{\sigma}^4}{\pi^2}
      \left\{8-\frac{3m_\sigma^2}{4 M^2}
              -\frac{9}{2}\int^1_0dxln\left[1-\frac{m_\sigma^2}{M^2}x(1-x)\right]\right\}.
\end{eqnarray}

\section{Numerical results}
\label{sec: num}

In Ref.~\cite{suzuki2}, the modification of the $\omega$ meson
mass in nuclear medium due to the vacuum polarization has been
studied within the Walecka model. They have taken the values of
the masses as, $M=939\ \rm{MeV}$,  $m_{\omega}=783\ {\rm MeV}$,
and $m_{\sigma}=520\ {\rm MeV}$. The coupling constants,
$g^2_\sigma=66.117$ and $g^2_\omega=79.927$, were determined by
requiring that the renormalized Hartree approximation could
reproduce the binding energy $-15.75 \ {\rm MeV}$ and the
equilibrium Fermi momentum $1.3 \ {\rm fm^{-1}}$ of nuclear
matter. Here we use the same parameters to calculate the effective
masses of $\sigma$ and $\omega$ mesons in nuclear medium, which
can be obtained by searching for the zeros of the inverse
propagators given in Eqs. (\ref{eq:msw}) and (\ref{eq:mww}). We
show in Fig. 1 the effective masses of the nucleon, $\sigma$ and
$\omega$ mesons as a function of the density. One can see that
both $M^*$ and $m^*_\omega$ decrease at finite density in the
Walecka model, which are in agreement with the results in
Ref.~\cite{suzuki2}. The effective mass of $\sigma$ meson,
$m^*_\sigma$, decreases at lower densities, and then slightly
increases at higher densities. These results will be compared
with those obtained in the chiral sigma model.

\begin{figure}
\epsfxsize = 12 cm
\centerline{
\includegraphics[width=0.8\textwidth]{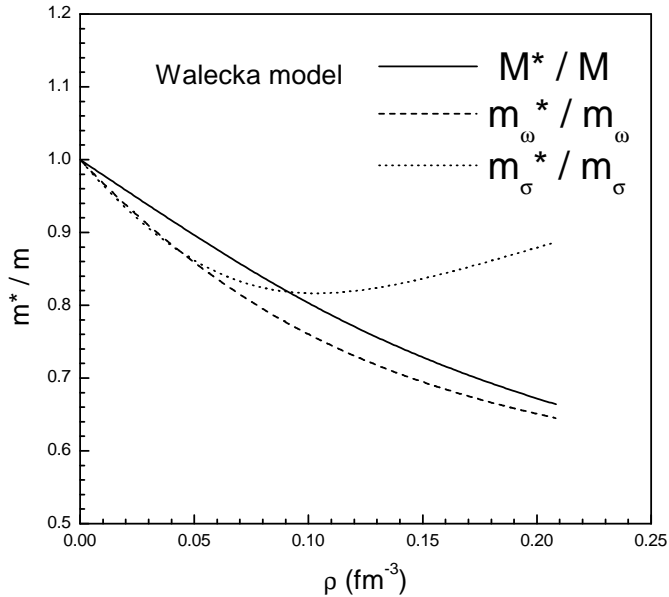}}
\vspace*{-6.5cm}
\caption{The effective masses of the nucleon, $\sigma$ and $\omega$ mesons
as a function of the baryon density within the Walecka model.}
\end{figure}

\begin{figure}
\epsfxsize = 12 cm
\centerline{
\includegraphics[width=0.8\textwidth]{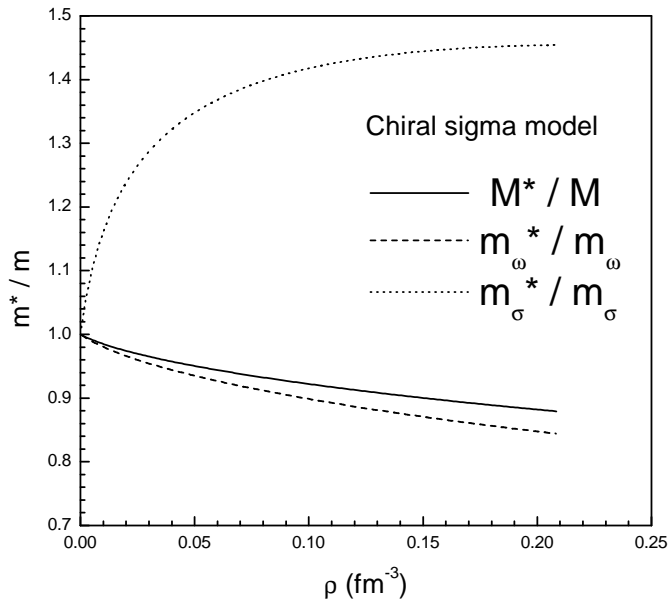}}
\vspace*{-6.5cm}
\caption{The effective masses of the nucleon, $\sigma$ and $\omega$ mesons
as a function of the baryon density within the chiral sigma model.}
\end{figure}

In the chiral sigma model, we take the vacuum expectation value of the $\sigma$ field as
the pion decay constant, $\sigma_0=f_{\pi}=93 \ {\rm MeV}$.
We adopt the masses $M = 939\ {\rm MeV}$, $m_{\pi} = 139\ {\rm MeV}$,
and $m_{\omega} = 783\ {\rm MeV}$ from their experimental values.
Then, the other parameters can be fixed automatically by the
following relations, $g_{\sigma}  = \frac{M}{f_\pi}  = 10.1$ and
$\widetilde{g}_{\omega} = \frac{m_{\omega}}{f_{\pi}} = 8.42$.
The coupling constants $\mu$ and $\lambda$ depend on $m_\pi$ and $m_\sigma$
through the relations given in Eqs. (\ref{eq:ms1}) and (\ref{eq:ms2}).
In the present model, $m_\sigma$ and $g_\omega$ are taken as free parameters,
which can be determined by reproducing the binding energy $-15.75 \ {\rm MeV}$
and the equilibrium Fermi momentum $1.3 \ {\rm fm^{-1}}$ of nuclear matter.
The fitted values for these two parameters in the renormalized Hartree approximation
are $m_\sigma = 715\ {\rm MeV}$ and $g_\omega = 4.025$.

We now present the results for hadron masses in nuclear medium
using the chiral sigma model, which are obtained by finding the
zeros of the inverse propagators given in Eqs. (\ref{eq:msc}) and
(\ref{eq:mwc}). In Fig. 2, we plot the effective masses of the
nucleon, $\sigma$ and $\omega$ mesons as a function of the
density. We observe that the reductions of $M^*/M$ and
$m^*_\omega/m_\omega$ in the chiral sigma model are slower than
those in the Walecka model, but the relationship between these two
ratios is kept to be the same in the two models. It is because
that the relationship derived in the Walecka model is also valid
in the chiral sigma model, which does not depend on the special
models and its parameters. In the chiral sigma model, the main
feature is that there is not much arbitrariness in the
renormalization procedure in order to respect the chiral symmetry.
The counterterms with the coefficients $\alpha_3$ and $\alpha_4$
in the chiral sigma model give larger nonlinear contributions, so
that it leads to smaller mean field value of $\sigma$ meson, which
is equivalent to larger effective nucleon mass. The effective mass
of $\sigma$ meson shown in Fig. 2 increases at finite density in
contrast to the Walecka model case. It is again due to the large
nonlinear $\sigma$ meson interactions.

The existence of $\sigma$ meson has been a controversial subject
for many years. Recently, there are a large number of evidences
showing its existence~\cite{pdg,E791}. However, the nature of
$\sigma$ meson as a conventional $q\bar{q}$ state or as a $\pi\pi$
resonant state is still under debate. The density dependence of
the $\sigma$ meson properties has been studied both
experimentally~\cite{CHAOS,TAPS} and
theoretically~\cite{oset,hatsuda2,thomas,chanfray}.
The measurements of the in-medium $\pi\pi$ masses were obtained
by the two-pion production experiments induced
either by pions (CHAOS collaboration)~\cite{CHAOS}
or by photons (TAPS collaboration)~\cite{TAPS}
on various nuclei. A significant nuclear-mass dependence
of the $\pi\pi$ invariant mass distribution in the $I=J=0$ channel
was observed in the experiments, which could be interpreted
as a signature for an in-medium modification of the $\pi\pi$ interaction
in the $\sigma$ channel.
On the theoretical side, the density dependence of the
$\sigma$ mass has been discussed in several models.
Vacas et al.~\cite{oset} calculated the $\pi\pi$ interaction in the
$\sigma$ channel at finite densities in a chiral unitary approach,
and found a dropping of the $\sigma$ mass as a function of the density.
In Ref.~\cite{hatsuda2}, Hatsuda et al. studied the
$\sigma$ propagator and found a decrease of the $\sigma$ mass
caused by the partial restoration of chiral symmetry.
However, the $\sigma$ mass in medium was found to be almost constant
in Ref.~\cite{thomas} by using a hybrid model for nuclear matter,
in which the nucleon, described as a quark-diquark state using the
NJL model, could be moving in self-consistent scalar and vector fields.
In the present work, we study the medium modification of
hadron masses by using the Walecka model and the chiral sigma
model with the vacuum polarization effects taken into account.
These models could reproduce reasonably well the saturation
properties of nuclear matter and the ground state properties of
finite nuclei~\cite{walecka,ogawa}. Therefore, it is very
interesting to discuss the density dependence of the $\sigma$ mass
in these models. The effective mass of $\sigma$ meson in the
Walecka model decreases at lower densities, and then slightly
increases at higher densities as shown in Fig. 1. However, we
obtain the raise of the $\sigma$ mass in medium,
as shown in Fig. 2, in the chiral sigma model using the standard
method of the introduction of the counterterms~\cite{matsui}.
It is the consequence of the negative energy states.
This behavior should be dependent on the renormalization procedure,
in particular, how to introduce the counterterms.

\section{Summary}
\label{sec: sum}

We have studied the variation of hadron masses in nuclear matter
due to the vacuum polarization using the chiral sigma model, and
compared the results with those obtained in the Walecka model.
Because of the constraints from chiral symmetry, there is less
arbitrariness in the renormalization procedure for the chiral
sigma model, as compared with the Walecka model. The renormalized
chiral sigma model is able to provide proper binding energy and
equilibrium density of nuclear matter, but the $\sigma$ meson mean
field value comes out to be too small due to the large nonlinear
$\sigma$ meson interactions, and it leads to a small reduction of
nucleon mass in medium. The effective mass of $\omega$ meson
decreases in nuclear matter. The reduction of $\omega$ meson mass
in the chiral sigma model is slower than those in the Walecka
model, while the relationship between $M^*/M$ and
$m^*_\omega/m_\omega$ is kept to be the same in the two models.
For $\sigma$ meson, its effective mass in the chiral sigma model
increases at finite density, which is an opposite behavior to the
results obtained in the Walecka model. It is very interesting to
see that the vacuum polarization do play an important role in the
modification of hadron masses in nuclear medium.

\section*{Acknowledgments}
We acknowledge fruitful discussions with Y. Ogawa and A. Haga.
This work was supported in part by the NSFC under contract No. 10135030,
and by the EYTP of MOE of China.



\end{document}